\documentstyle[12pt,tighten,floats,aps,epsf]{revtex}

\newcommand{\He}{^3\mbox{He}}
\newcommand{\ggama}{{\mit \Gamma}}

\newcommand{\be}{\begin{equation}}
\newcommand{\ee}{\end{equation}}
\newcommand{\bphi}{\bar{\phi}}
\newcommand{\NNpiNN}{$N\!N\!-\pi N\!N$}
\newcommand{\NNN}{$N\!N\!N$}
\newcommand{\tU}{\tilde{U}}
\newcommand{\tZ}{\tilde{Z}}
\newcommand{\eqn}[1]{\label{#1}}
\newcommand{\eq}[1]{Eq.~(\ref{#1})}
\newcommand{\eqs}[1]{Eqs.~(\ref{#1})}
\newcommand{\fign}[1]{\label{#1}}
\newcommand{\fig}[1]{Fig.~\ref{#1}}

\newcommand{\AmS}{{\protect\the\textfont2
  A\kern-.1667em\lower.5ex\hbox{M}\kern-.125emS}}

\title{Gauging hadronic systems}

\author{A. N. Kvinikhidze\thanks{On leave from The Mathematical Institute of
Georgian Academy of Sciences, Tbilisi, Georgia.}
        and
        B. Blankleider\thanks{The authors would like to
thank the Australian Research Council for their financial support.}}
\address{Department of Physics, The Flinders University
        of South Australia, \\
        Bedford Park, SA 5042, Australia}
\begin{document}
\maketitle

\begin{abstract}
We present a general method for incorporating the electromagnetic interaction
into descriptions of hadronic processes based on four-dimensional scattering
integral equations. The method involves the idea of gauging the scattering
equations themselves, and results in electromagnetic amplitudes where an
external photon is effectively coupled to every part of every strong interaction
diagram in the model.  Current conservation is therefore implemented in the
theoretically correct fashion. To illustrate our gauging procedure we apply it
to the three-nucleon problem whose strong interactions are described by standard
three-body integral equations. In this way we obtain the expressions needed to
calculate all possible electromagnetic processes of the three-nucleon system:
the electromagnetic form factors of the three-body bound state, $pd\rightarrow
pd\gamma$, $\gamma ^3\mbox{He}\rightarrow pd$, $\gamma \He\rightarrow ppn$,
etc. As the photon is coupled everywhere in the strong interaction model, a
unified description of the \NNN -$\gamma$\NNN\ system is obtained. An
interesting aspect of our results is the natural appearance of a subtraction
term needed to avoid the overcounting of diagrams.
\end{abstract}

\section{INTRODUCTION}
There is currently great interest in the use of photons and electrons to probe
the structure of hadronic systems. As the strong interactions of hadrons are
often described through the solution of integral equations, there is an
outstanding problem of how to extend such non-perturbative descriptions to the
electromagnetic sector. An essential requirement of such an extension is that
of  electromagnetic current conservation.

Here we would like to present a simple but general method for constructing
conserved currents that is applicable to any hadronic process defined by
relativistic four-dimensional scattering equations. The method involves a direct
gauging of the equations themselves and results in the photon being effectively
coupled everywhere in the strong interaction model, so that current conservation
is guaranteed.

To illustrate our gauging procedure, we first apply it to the
two-body problem, and then to the considerably more complicated case of the
relativistic three-body problem. In gauging the three-body case there exists
an overcounting problem which has previously been overlooked. It is a feature
of our gauging method that it solves this overcounting problem in an automatic
way. We then apply our results to the three-nucleon system where the further
requirement of antisymmetry is fully taken into account.

Although the power of the gauging of equations method is thus well demonstrated
on the example of the three-nucleon system, it should be emphasised that the
same method is easily applied to systems where the number of hadrons is not
conserved. Indeed we have recently used this method to gauge the the \NNpiNN\
system \cite{gpinn} which has overcounting problems all of its own even before
the attachment of photons. The same method can also be used to generate
three-dimensional descriptions of gauged hadronic systems by applying it to the
corresponding spectator equations \cite{KB1_97}. In this respect we have
recently solved the long-standing problem of how to describe the gauged
three-nucleon system within the spectator approach \cite{KB2_97}.

Finally, it is important to note that although we are concerned in this paper
with the electromagnetic interaction for which current conservation is a major
issue, the gauging of equations method itself is totally independent of the type
of external field involved. Thus the results of this paper hold also for the
case of a weak probe for which current is not conserved (of course the gauged
inputs would need to be chosen appropriately).

\section{GAUGING TWO PARTICLES}

\subsection{Gauging the two-body Green function}
We consider the scattering of two distinguishable hadrons described by the Green
function $G$. Then the corresponding quantity with a ``photon attached'' is
given by the five-point function $G^\mu$ (see for example Ref.\cite{Bentz}). If
$G$ and $G^\mu$ are exact, as specified by their definitions in field theory,
then they satisfy the Ward-Takahashi (WT) identity \cite{WT,Bentz}
from which current conservation follows.

Here we demonstrate how to find $G^\mu$ when $G$ is given only within a
particular model. Moreover, we show how to attach photons everywhere to $G$ so
that the WT identity is still satisfied. We start by expressing $G$ in terms of
its fully disconnected part, $G_0$, and the model-dependent kernel $V$:
\be G=G_0+G_0VG . \eqn{G}
\ee
This is a symbolic equation representing, in the case of two-particle
scattering, a four-dimensional integral equation.  \eq{G} is basically a
topological statement regarding the two-particle irreducible structure of
Feynman diagrams belonging to $G$; as such, it can be utilised directly to
express the structure of the same Feynman diagrams, but with photons attached
everywhere. Thus from \eq{G} it immediately follows that
\be
G^\mu =G_0^\mu + G_0^\mu VG+G_0V^\mu G+G_0VG^\mu .  \eqn{G^mu}
\ee
This result expresses $G^\mu$ in terms of an integral equation, and illustrates
what we mean by {\em gauging an equation}, in this case the gauging of \eq{G}.
Implied in \eq{G^mu} is the result
\be
[G_0VG]^\mu = G_0^\mu VG+G_0V^\mu G+G_0VG^\mu ,  \eqn{product}
\ee
which illustrates a rule for the gauging of products that is rather reminiscent
of the product rule for derivatives. Although \eq{product} follows from a
topological argument, we shall postulate such a product rule to hold also in
other cases where topological arguments may not be applicable. In \eq{G^mu},
both $G^\mu$ and $G_0^\mu$ are obtained from the Green functions $G$ and $G_0$,
respectively, by attaching photons everywhere. The gauged potential $V^\mu$ is
similarly obtained from $V$, but with no photons attached to external legs since
such contributions are already taken into account in the terms $G_0^\mu VG$ and
$G_0VG^\mu$. \eq{G^mu} is now easily solved algebraically giving
\be
G^\mu =G\Gamma ^\mu G,         \eqn{Gamma^mudef}
\ee
with the electromagnetic vertex function $\Gamma^\mu$ being given by
\be
\Gamma^\mu =\Gamma_0^\mu+V^\mu \hspace{1cm};\hspace{1cm}
\Gamma_0^\mu=G_0^{-1}G_0^\mu G_0^{-1}.
\eqn{Gamma^mu}
\ee
For a two-particle system $G_0=d_1d_2$ so that $\Gamma_0^\mu = \ggama_1^\mu
d_2^{-1} + d_1^{-1} \ggama_2^\mu \eqn{Gamma_0^mu}$ where $d_i$ ($i=$1,2) is the
Feynman propagator of particle $i$ and $\ggama_i^\mu$ is the one-particle
electromagnetic vertex function defined by $d_i^\mu = d_i\, \ggama_i^\mu d_i$.
$\Gamma_0^\mu$ is thus the sum of single nucleon currents, and $V^\mu$ is the
two-body interaction current of the given model. For the case of two particles,
this result agrees with the one of Gross and Riska \cite{GR} who used a
different method, that of analysing Feynman diagram contributions, to derive
this result. It is easy to show explicitly that the $G^\mu$ as specified by
\eqs{Gamma^mudef} and (\ref{Gamma^mu}) satisfies the WT identity.

\subsection{Gauging the two-body bound state wave function}
So far we have defined ``gauging'' to be the process where photons are
attached to all places in perturbation diagrams. As Green functions and
potentials have a diagrammatic interpretation, the gauging of these quantities
has therefore a clear meaning.  On the other hand, the bound state wave function
is a purely nonperturbative quantity, and thus cannot be associated with
diagrams. Nevertheless, we can define the gauged bound state wave
function by formally gauging the bound-state Bethe-Salpeter equation using our
product rule. Thus the gauging of the bound state equation
$
\psi = G_{0}V\psi \eqn{psi}
$
is postulated to give
$
\psi^\mu = \left(G_{0}V\right)^\mu\psi+G_{0}V\psi^\mu. 
$ Simple algebra then gives the explicit expression for $\psi^\mu$:
\be
\psi^\mu = G \Gamma^\mu \psi.   \eqn{psi^mu_2}
\ee
As \eq{psi^mu_2} can also be obtained from the $G^\mu$ of \eq{Gamma^mudef} by
taking the right hand residue at the two-body bound state pole, we see that
$G_0^{-1}\psi^\mu$ is just the transition amplitude for the
photo/electro-disintegration of the hadronic bound state. The bound state vertex
function $\phi$, defined by $\psi=G_0\phi$, may be gauged in the same way.
Interestingly, in contrast to $G_0^{-1}\psi^\mu$ on mass shell, neither
$\psi^\mu$ nor $\phi^\mu$ conserve current.

\section{GAUGING THREE PARTICLES}
\subsection{Gauging the three-body Green function}
It is clear that the derivation of \eqs{Gamma^mudef} and (\ref{Gamma^mu}) does
not depend on the number of particles. It therefore applies when $G$ is the
Green function for three particles and $G^\mu$ is the corresponding seven-point
function.  In this case however $\Gamma_0^\mu=\sum_{i=1}^3\ggama_i^\mu
D_{0i}^{-1}$, $D_{0i}=d_jd_k$,
and $V^\mu$ refers to the gauged three-body potential. In the absence of
three-body forces, $V$ is a sum of three disconnected potentials:
$V=V_1 + V_2 + V_3$.
Here we use the usual spectator notation where $V_i$ is the potential describing
the interaction of particles $j$ and $k$ with particle $i$ being a spectator
($ijk$ is a cyclic permutation of $123$). Explicitly we
have that
\be
V_i = v_i d_i^{-1}   \eqn{V_i_short}
\ee
where $v_i$ is the two-body potential.
As $V^\mu=V_1^\mu + V_2^\mu + V_3^\mu$, to obtain $V^\mu$ it is sufficient to
gauge \eq{V_i_short}. To gauge the inverse propagator $d^{-1}$ we formally gauge
the identity $d^{-1}d=1$, in this way obtaining
$\left(d^{-1}\right)^\mu =-d^{-1}d^\mu d^{-1} = -\ggama^\mu.$ Thus
$
V_i^\mu = v_i^\mu d_i^{-1}-v_i \ggama_i^\mu.
$
The negative sign in front of the term $v_i \ggama_i^\mu$ may appear to be
surprising, yet it is just what is needed to stop overcounting. Consider, for
example, the gauging of the term $G_0VG$ appearing in \eq{G}:
$(G_0VG)^\mu=G_0^\mu VG+G_0V^\mu G+G_0VG^\mu$. It is apparent that
$v_i \ggama_i^\mu$
appears in each of the three terms $G_0^\mu VG$, $G_0V^\mu
G$, and $G_0VG^\mu$. Thus the negative sign in question is needed to ensure that
$v_i \ggama_i^\mu$ contributes only once to the gauging of $G_0VG$.
The electromagnetic vertex function for the three-particle system is therefore
\be
\Gamma^\mu=\sum_{i=1}^3 \left(\ggama_i^\mu D_{0i}^{-1}+v_i^\mu d_i^{-1}
-v_i\ggama_i^\mu \right)  \eqn{hello}
\ee
which we illustrate in \fig{gamma_3d}.
\begin{figure}[t]
\hspace*{1cm}  \epsfxsize=14cm\epsfbox{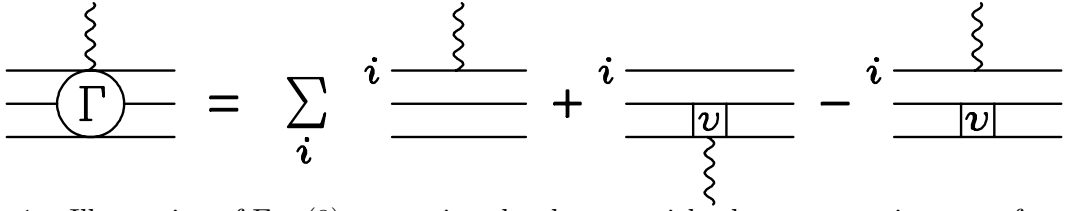}
\caption{\fign{gamma_3d} Illustration of \protect\eq{hello} expressing the
three-particle electromagnetic vertex function $\Gamma^\mu$ as a sum of one- and
two-nucleon currents.}
\end{figure}
It is interesting to note that had one written the expression for $\Gamma^\mu$
by analogy to what is done in standard non-relativistic theory, one would get
the wrong sign for $v_i\ggama_i^\mu$. Just this was done in the pioneering
four-dimensional calculations of the \NNN\ system by Rupp and Tjon \cite{RT}; as
a result, this work contains overcounting.

\subsection{Gauging the AGS equations}
Although one can obtain the electromagnetic transition amplitudes for {\em any}
process by taking appropriate bound state residues of $G^\mu$, for the
three-body problem there exists an alternative approach that is especially
convenient. This approach is based on the four-dimensional version of the
Alt-Grassberger-Sandhas (AGS) equations \cite{AGS} which describe the scattering
of three strongly interacting hadrons. The AGS equations provide a direct way to
calculate amplitudes for physical processes like $N\!-d$ elastic scattering or
breakup. They also take as input two-body t-matrices $t_i$ rather than the
two-body potentials $v_i$ which form the input of \eq{G} for the Green function
$G$. Thus by gauging the AGS equations one obtains electromagnetic transition
amplitudes in a direct way, with the further advantage of having gauged
t-matrices $t_i^\mu$ as input rather than gauged potentials $v_i^\mu$ which form
the input of $G^\mu$.

For distinguishable particles the AGS equations can be written in the matrix
form
\be
\tilde{U}={\cal I}G_0+{\cal I}G_0T\tilde{U}.    \eqn{tildeU}
\ee
Here $\tU$ is a matrix whose the $(i,j)$'th element is defined by
$\tU_{ij}=G_0U_{ij}G_0$ where $U_{ij}$ is the usual AGS amplitude describing
the the scattering of particle $j$ off the $(ki)$ quasi-particle leading to a
final state consisting of particle $i$ and the $(jk)$ quasi-particle. The other
matrix quantities in \eq{tildeU} are defined by
$
T_{ik} = \delta_{ik}t_kd_k^{-1}
$
and
$
{\cal I}_{ik}  = 1-\delta_{ik}.
$
\eq{tildeU} may now be gauged in just the same way as \eq{G}. We obtain that
\be
\tilde{U}^\mu=\tilde{U}\Gamma ^\mu\tilde{U}     \eqn{tU^mu}
\ee
where $\Gamma^\mu$ is a matrix whose $(n,m)$'th element is
\be
\Gamma^\mu_{nm}
= {\cal I}^{-1}_{nm} \sum_{i=1}^3  \ggama_i^\mu D_{0i}^{-1}
+ \delta_{nm} t_n^\mu d_n^{-1} - \delta_{nm} t_n \ggama_n^\mu . \eqn{Umu}
\ee
This result can now be used to calculate electromagnetic processes of three-body
systems. For example, to calculate $j(ki)\rightarrow i(jk)\gamma$ where $(ki)$
and $(jk)$ represent two-body bound states with bound state vertex functions
$\phi_j$ and $\phi_i$ respectively, we first write the physical t-matrix for the
process $j(ki)\rightarrow i(jk)$ as $T_{ij} = d_i^{-1} \bphi_i \tU_{ij} \phi_j
d_j^{-1}$, and then gauge $T_{ij}$ using the product rule.

\subsection{Application to the gauged three-nucleon system}

For identical particles there is a variety of ways to
define the AGS amplitude, all giving the same three-body Green function $G$.
The one we have chosen, which we call $Z$, satisfies the particularly simple 
AGS equation for identical particles
\be
\tZ = G_0 + D_{03}t_3{\cal P}\tZ   \eqn{Z}
\ee
where $\tZ=G_0ZG_0$ and where ${\cal P}$ shifts particle labels cyclically to
the left.  As $\tZ$ is related to the three-body Green function $G$ in a simple
way, the solution to our problem of specifying $G^\mu$ now rests essentially on
the construction of the gauged AGS Green function $\tZ^\mu$. Gauging \eq{Z} in
the same way we gauged \eq{G}, we obtain \be \tZ^\mu=\tZ d_3^{-1}d_3^\mu
+\tZ\left( D_{03}^{-1}D_{03}^\mu D_{03}^{-1}d_3^{-1} +d_3^{-1}t_3^\mu{\cal
P}\right) \tZ .  \eqn{Zmu} \ee This equation describes the attachment of photons
at all possible places in the multiple-scattering series of three identical
particles. The input is in terms of two-body t-matrices. As such, it forms the
central result in the gauged three-nucleon problem.

\end{document}